\documentclass[prd,aps, reprint, nolongbibliography]{revtex4-2}
\usepackage{tikz}
\usepackage{graphicx}
\usepackage{amsmath}
\usepackage{amssymb}
\usepackage{multirow}

\newcommand\Pxgb{$P_{\mathrm{XGB}}$ }

\usepackage[citecolor=blue,linkcolor=blue,colorlinks=true,urlcolor=blue]{hyperref}

\usepackage{float}
\usepackage{hyperref}

\begin{document}

%\title{Enhancing detection of gravitational waves with machine learning}
\title{Optimization of model independent gravitational wave search using machine learning}
\author{Tanmaya Mishra, Brendan O'Brien, V. Gayathri, Marek Szczepa\'nczyk, Shubhagata Bhaumik, Imre Bartos and Sergey Klimenko}
\affiliation{University of Florida}

\begin{abstract}
 The Coherent WaveBurst (cWB) search algorithm identifies generic gravitational wave (GW) signals in the LIGO-Virgo strain data.
We propose a machine learning (ML) method to optimize the pipeline sensitivity to the special class of GW signals known as binary black hole (BBH) mergers.
%XGBoost provides robust classification between GW signals and noise transients which improves significance of GW candidates, thereby supplementing the cWB search.
Here, we test the ML-enhanced cWB search on strain data from the first and second observing runs of Advanced LIGO and successfully recover all BBH events previously reported by cWB, with higher significance. For simulated events found with a false alarm rate less than $1\,\mathrm{yr}^{-1}$, we demonstrate the improvement in the detection efficiency of $26\%$ for stellar-mass BBH mergers and $16\%$ for intermediate mass black hole binary mergers. To demonstrate the robustness of the ML-enhanced search for the detection of generic BBH signals,  we show that it has the increased sensitivity to the spin precessing or eccentric BBH events, even when trained on simulated quasi-circular BBH events with aligned spins.
%To solve this problem, we propose a machine learning method to automate the selection of the cWB events and to improve the performance of the cWB algorithm for detection of the binary black hole (BBH) mergers. We use XGBoost, a decision tree based ensemble learning algorithm, to provide a ranking statistic for efficient separation of signal and noise. Using the XGBoost statistic we construct a new cWB detection statistic for more efficient detection of BBH events, which would have been otherwise missed with the standard cWB selection cuts. We incorporate XGBoost techniques into the cWB framework and test it on the data from the first and second observing runs of advanced LIGO-Virgo (O1 and O2). We demonstrate an improvement of \textcolor{red}{$\sim 25\%$} of the detection efficiency on a simulation set of stellar-mass BBHs, and an improvement of \textcolor{red}{$\sim 15\%$} for intermediate mass black hole mergers. In this analysis, we recover with higher significance all of the BBH candidate events previously detected by cWB and also recover other potential BBH events with the false-alarm rate (FAR) $< 1 \, \mathrm{yr}^{-1}$ previously missed by the cWB search. Finally, we demonstrate the robustness of the algorithm for detection of generic BBH mergers by testing on a set of simulated eccentric BBH waveforms.

\end{abstract}

\maketitle

%%%%%%%%%%%%%%%%%%%%%%%%%%%%%%%%%%%%%%%%%%%%%%%%
\section{Introduction\label{intro:sec}}

The detection of the first gravitational wave (GW) signal GW150914~\cite{GW150914} commenced the age of the GW astronomy.
Since then, the Advanced LIGO~\cite{TheLIGOScientific:2014jea} and Advanced Virgo~\cite{TheVirgo:2014hva} network of detectors have identified 11 GW candidates during the first two observing runs (O1 and O2)~\cite{GWTC1}, 40 GW candidates in the first half of the third observing run (O3a)~\cite{GWTC2}, and provided 20 GW public alerts to EM astronomers during the second half of the third observing run (O3b)~\cite{GraceDB-03-Public}.
With the improving sensitivity of the GW detector network, it is essential to refine the search algorithms used to detect GW signals.  

Coherent WaveBurst (cWB) is an algorithm that searches for excess power in the time-frequency domain to identify GW signals in the LIGO-Virgo strain data~\cite{Klimenko:2008fu, Klimenko:2016}.
Unlike other analysis pipelines which search for binary black hole (BBH) mergers, cWB does not use template waveform models.
Instead, the cWB algorithm is model independent, which makes it a valuable tool in the search for poorly modeled or unexpected GW sources.
cWB played an integral role in the discovery of the first BBH merger GW150914~\cite{GW150914} and, more recently, in the first direct detection of an intermediate mass black hole (IMBH) GW190521~\cite{GW190521.1-Discovery, cwb_GW190521}. Also, cWB has contributed to the detection of 22 BBH events in the O1, O2, and O3a observing runs~\cite{GWTC1, GWTC2}.

The cWB pipeline generates summary statistics for every identified event. These summary statistics
describe generic properties of reconstructed events such as the characteristic frequency, duration, cross-correlation between different detectors, and other parameters described in Appendix A. 
%describe the signal-like or noise-like properties for a given event.
In the standard cWB search framework, we use summary statistics to construct vetoes designed to reject noise events from the analysis and increase the significance of detected GW events.
%, which are used to separate signal events from background noise events.
%To improve the detection efficiency, standard vetoes are constructed by making use of the various summary statistics estimated by the pipeline. 
%Events which do not pass these vetoes are omitted from the analysis.
Although this veto procedure generally works well, it risks the removal of GW signal events that lie near the border of the predefined veto thresholds.
In addition, designing vetoes is challenging since they need to be redefined for each detector network and are dependent on the run conditions.
%Generally, this veto procedure works well but risks complete removal of GW signal events which sit on the borderline of the vetoes. 

Machine learning (ML) offers a novel approach to solving complex problems. Accordingly, the interest of ML techniques in GW astronomy has grown in recent years~\cite{Cuoco_2020} as ML has been applied to categorize noise artifacts in the GW detector strain data~\cite{PhysRevD.101.102003, Zevin_2017, wavelet_XGB}, classify GW signals~\cite{PhysRevD.100.063015, Iess_2020}, and estimate GW source parameters~\cite{George2018, Schmidt_2021}.
ML has already been used in combination with cWB for various other studies~\cite{Vinciguerra_2017, Cavagli__2020, Gayathri_2020, OBrien_2021}.

In this paper, we propose to use a decision tree based ensemble learning algorithm called eXtreme-Gradient Boost (XGBoost)~\cite{XGBoost} to automate the signal-noise classification in cWB and optimize the pipeline sensitivity to BBH mergers.
%We do not use ML as an independent GW detection method but instead use ML to enhance the existing cWB architecture.
To preserve the waveform independent analysis, we do not attempt to train the ML model directly on the GW strain data. Instead, we utilize cWB to reconstruct events and generate their summary statistics, and then we carefully select a subset of summary statistics used for the construction of the ML model.
The end result is the ML-enhanced search pipeline which is resistant to overfitting and provides the robust recovery of GW events with the waveform parameters that could be outside of the training set.
We test the cWB pipeline enhanced with XGBoost on publicly available LIGO Hanford and LIGO Livingston strain data from O1 and O2~\cite{Open_data_O1_O2}.

%Here, we do not use ML as an independent GW detection method but as an enhancement to the existing cWB detection algorithm. cWB reduces the dimensionality of the problem by mapping information, generic to binary black holes, into summary statistics.
%We feed the cWB summary statistics into XGBoost Classifier and use the output to construct the new veto statistic. We redefine the cWB detection statistic by combining this ML enabled veto statistic with the standard cWB detection statistic.
%The detection efficiency reported in this paper is calculated with respect to the total number of recovered events by the pipeline, defined in Section~\ref{res2:sec}.

%We demonstrate that our detection efficiency to stellar-mass BBH mergers improves by \textcolor{red}{$26\%$}, and our detection efficiency to IMBH binary mergers improves by \textcolor{red}{$16\%$}.
%We recover all BBH events previously detected by the standard cWB search in addition to GW170809, which was previously missed by the search.
%We also show that the enhanced cWB pipeline is able to detect eccentric BBHs at a greater sensitivity than the standard search.

% We do not identify any new BBH event candidates with a false-alarm rate (FAR) $< 1$ per year.

The paper is organized as follows. In Section~\ref{cWB:sec}, we introduce the cWB search pipeline.
In Section~\ref{data:sec}, we describe the data used to train and test our ML algorithm.
In Section~\ref{ML:sec}, we demonstrate the implementation of ML into the detection procedure and define our new cWB reduced detection statistic.
In Section~\ref{res:sec}, we compare the sensitivity of the ML-enhanced cWB search against the sensitivity of the standard cWB search. We report the updated significance of BBH events detected during the O1 and O2 observing runs.
Finally, in Section~\ref{conclusion:sec}, we state the conclusions of our study.

%%%%%%%%%%%%%%%%%%%%%%%%%%%%%%%%%%%%%%%%%%%%%%%%
\section{cWB search algorithm\label{cWB:sec}}

The cWB search algorithm is designed to detect GW signals with minimal assumptions on the signal model~\cite{Klimenko2008, Klimenko:2016}.
The detector strain data is mapped to the time-frequency domain using the Wilson Daubechies Meyer (WDM) wavelet transform~\cite{Necula:2012zz} where the data is normalized by the rms of the detector noise.
The algorithm then identifies WDM wavelets with excess power above the average fluctuations of the detector noise. The selected nearby wavelets are grouped into clusters.
%in blocks of fixed time-frequency area called \textit{pixels}.
%Pixels with power above the threshold for detector noise are collected and grouped into clusters.
The pipeline generates an event for each selected cluster and reconstructs the signal waveform using the constrained maximum likelihood method~\cite{Klimenko:2016}.
%-pixels: fixed time-frequency area
%-The algorithm selects pixels which are above the threshold for detector noise fluctuations. 
%-generates an event
%-reconstructed the waveform using max likelihood procedure

%The cWB pipeline generates the reconstructed signal waveforms for both the GW events as well as the non stationary noise transients and instrumental glitches. 
For each event, various summary statistics are estimated by the pipeline which describes the time-frequency structure, signal strength, and coherence across the detector network.
%The coherent energy $E_c$ is obtained by cross-correlating the reconstructed signal waveforms across different detectors, the residual noise energy $E_n$ estimated by subtracting the reconstructed waveforms from the data.
The main detection statistic for the cWB generic GW search is the signal-to-noise ratio (SNR)  defined for the LIGO detector network as:
\begin{equation} \label{eq:1}
\eta_\mathrm{0} = \sqrt{\frac{E_\mathrm{c}}{2\,\text{max}\left(1,\chi^2\right)}} \,. \end{equation}
Here, $E_\mathrm{c}$ denotes the coherent energy estimated by cross-correlating the reconstructed signal waveforms across different detectors, and $\chi^2 = E_\mathrm{n} / N_{\mathrm{df}}$ where $E_\mathrm{n}$ is the estimated residual noise energy and $N_{\mathrm{df}}$ is the the number of independent wavelet amplitudes describing the event. The $\chi^2$ correction in Equation~\ref{eq:1}, which is close to unity for genuine GW events, reduces the non-Gaussian noise contribution.
For the cWB searches which target BBH events, the detection statistic is modified to favor events which frequency is increasing with time:
\begin{equation}\label{eqn2}
\eta_\mathrm{1} = \eta_\mathrm{0} \, F_{\mathrm{M}} \, \sqrt{e_{\mathrm{M}}} \, ,
\end{equation}
%where $W_{\mathrm{M}}$ is the chirp mass penalty factor and is given by $W_{\mathrm{M}} = F_{\mathrm{M}} \, \sqrt{e_{\mathrm{M}}}$. Here, 
where $F_{\mathrm{M}}$ is the event energy fraction and $e_{M}$ is the event ellipticity defined in Ref.~\cite{Tiwari2015_chirp}. Both parameters  $F_{\mathrm{M}}$ and $e_{M}$ are close to unity for BBH events and penalize events which time-frequency evolution is significantly different from the chirping BBH signal. 

GW detector data is hindered by noise artifacts known as glitches, and consequently, some noise events are reconstructed by the pipeline and leak into the analysis. In the standard cWB analysis, we apply a series of vetoes to target and remove these glitches. 
This approach, henceforth known as the \textit{veto method}, improves the significance of candidate GW events by reducing excess background. 
The veto method consists of applying a priori defined veto thresholds on a set of the cWB summary statistics. This procedure discretely classifies generated events into one of the two categories: signal-like events and noise-like events. Events that fall into the noise-like category are removed from the analysis.
This process could inevitably result in discarding borderline GW events which do not pass the veto thresholds and at the same time makes the pipeline vulnerable to the high SNR glitches which do pass the vetoes. Designing vetoes in the multidimensional space of the summary statistics is challenging, and furthermore, requires re-tuning of the veto thresholds for each detector network configuration and each observing run.

%The cWB pipeline estimates various signal properties which % It gives this set of parameters for each trigger event --- a special time instance in the data that could potentially be a GW signal.
%The production stage of cWB generates the interesting trigger events, where as, the process of selecting the cuts is a part of the cWB post-production (PP) analysis. 

In the standard cWB setup, the veto method is tuned separately to improve the search sensitivity to stellar-mass BBH mergers ($M_{\mathrm{tot}} \lesssim 100\, M_\odot$) and IMBH binary mergers ($M_{\mathrm{tot}} \gtrsim 100\, M_\odot$).
%Although stellar-mass BBH mergers ($M_{\mathrm{tot}} \lesssim 100\, M_\odot$) and intermediate mass black hole (IMBH) binary mergers ($M_{\mathrm{tot}} \gtrsim 100\, M_\odot$)
While the GW waveforms of these two classes are conceptually similar, the corresponding GW signals observed in the LIGO frequency band are quite different. A GW signal originating from the stellar-mass BBH merger usually  exhibits the full inspiral-merger-ringdown waveform, whereas GW signals from IMBH binary mergers are short in duration and contain mostly the merger-ringdown waveform, with the inspiral signal buried inside the low-frequency seismic noise.
As a result, we utilize two configurations of the cWB search tuned for these systems: the BBH configuration which targets stellar-mass BBH mergers, and the IMBH configuration which targets IMBH binary mergers~\cite{cwb_GW190521}. IMBH binaries are expected to merge at lower frequencies compared to the stellar-mass BBH mergers, and so the corresponding cWB search configurations apply different vetoes to account for the difference in the signal morphology.

%In this paper, we examine each search configuration independently.

\section{Data} 
\label{data:sec}

We analyze publicly available strain data from Advanced LIGO's first two observational runs~\cite{Open_data_O1_O2}. Here, we only examine data from the LIGO Hanford and LIGO Livingston detectors, with the inclusion of Virgo data left for future work.
To train and test our ML model, we require a representative set of noise events and signal events.

% Noise events, also known as \textit{background} events, 
Noise events (background) are generated by systematically time-shifting the data from one detector with respect to other detectors in the detector network. Each time shift is chosen to be greater than the time of flight between detectors to exclude true astrophysical signals. This process is repeated multiple times for various time lags, and we count the number of background events generated over the total accumulated background time.
For the O1 run, we accumulated approximately 16,000 years and 4,000 years of the background time for the BBH and IMBH searches, respectively.
For the O2 run, we accumulated approximately 11,000 years of background for each search.

To generate a representative set of the signal events, we add (inject) simulated GW signals to the detector data and reconstruct them with cWB.
In this work, we investigate four sets of simulated signals: (i) a quasi-circular spin-aligned stellar-mass BBH set, (ii) a quasi-circular IMBH binary set, (iii) an eccentric BBH set, and (iv) a quasi-circular precessing BBH set. Only the first two simulation sets of the quasi-circular signals were used for ML training, whereas the remaining two sets are used to test the robustness of the ML implementation. In all four cases, the binary orientation parameters (sky location, inclination angle) for every simulated waveform are randomly drawn from uniform distributions. The redshift $z$ is drawn from a uniform distribution in co-moving volume, assuming Planck 2015 cosmology~\cite{Planck2015}.

To simulate the stellar-mass BBH set, we use the SEOBNRv3~\cite{Babak:2016tgq} and SEOBNRv4~\cite{Boh2017} waveform approximants.
These waveform approximants include only the dominant ($\ell = 2, m = 2$) harmonic mode.
The source frame total mass for these simulations ranges from approximately $5\,M_\odot$ to $100 \, M_\odot$, and the mass ratio $q = m_2/m_1$ ranges from approximately 1/4 to 1.
Component black hole spins are aligned with the orbital plane.
%-some spins drawn from aligned distribution, some spins drawn from isotropic distribution

For the IMBH binary set, we use numerical relativity waveforms which include higher-order harmonics. We consider 17 massbins, as used in Ref.~\cite{Abbott:2019ovz}, which range in source frame total mass from $120\, M_\odot$ to $800 \, M_\odot$, with mass ratios ranging from 1 to 1/10.

For the high mass, eccentric BBH set, we also use numerical relativity waveforms~\cite{Healy2017, gayathri2020gw190521}. We consider 28 massbins which range in total mass from $100 \, M_\odot$ to $250 \, M_\odot$, mass ratio equal to 1, with eccentricities ranging from 0.66 to 0.99.

For the precessing stellar-mass BBH set, we use the SEOBNRv4PHM~\cite{SEOBNRV4PHM} waveform approximant, which includes precession and higher-order harmonic modes.
The source frame total mass ranges from $4\, M_\odot$ to $200 \, M_\odot$, with mass ratios ranging from 1 to 1/20.
Component black hole spins are isotropically distributed.

%%%%%%%%%%%%%%%%%%%%%%%%%%%%%%%%%%%%%%%%%%%%%%%%
\section{Machine Learning implementation\label{ML:sec}}

%. The probability value of each event being a signal event is provided by the XGBoost Classifier. 
%A monotonic transformation of this XGBoost probability gives us a continuous ranking criteria, 
%which allows us to detect GW events which would have been otherwise missed with the standard PP selection cuts on cWB parameters.

The veto method used in the standard cWB search categorizes events into two discrete bins: signal events and noise events.
Here, the veto method effectively acts as a decision tree with only two leaves, where the summary statistics for a given event are compared against various rules at each decision node until the event is classified as either a signal event or a noise event.
Since this method produces a discrete outcome, it unavoidably removes signal events which do not pass all veto thresholds.

We propose using ML, which produces continuous ranking criteria for all reconstructed events, to replace the veto method.
Binary classification is a standard problem in the ML literature.
Moreover, many prominent ML algorithms are based on the decision tree structure, which we expect is well suited for the cWB classification problem.

We use the boosted decision tree based ML algorithm called XGBoost~\cite{XGBoost}.
In XGBoost, instead of using a single decision tree to classify events, an ensemble of decision trees is generated. A decision tree is used as the base learner, and subsequent learners (trees) are formed based on the residual errors obtained after each iteration (boosting).
This process is expected to be more accurate and more robust than using a single decision tree used by the veto method.
A continuous score is calculated by taking the weighted average of output, obtained from each decision tree in the ensemble. The final output \Pxgb is computed by taking the sigmoid of the continuous score, where a value close to zero denotes a noise-like event and a value close to one denotes a signal-like event.
%Although the value of \Pxgb ranges from zero to one, the interpretation of this quantity as a probability can be misleading.

To construct the ML model, we select a subset of 14 summary statistics estimated by cWB as input features for the ML algorithm.
Selected summary statistics describe the signal strength and the correlation across the detectors ($\eta_\mathrm{0}$, $c_\mathrm{c}$, $n_\mathrm{f}$), the quality of the likelihood fit ($E_\mathrm{c}/L$, $\chi^2$), the time-frequency evolution of the event ($\Delta T_\mathrm{s}$, $\Delta F_\mathrm{s}$, $f_0$, $\mathcal{M}$, $e_\mathrm{M}$) and the two different estimators for the number of cycles in the reconstructed waveform ($Q_\mathrm{a}$, $Q_\mathrm{p}$).
%While others still are expected to penalize noise-like events ($Q_\mathrm{a}$, $Q_\mathrm{p}$,).
%In our testing, we found that adding supplementary statistics contributed mostly redundant information and did not markedly improve classification.
%On the other hand, further feature pruning led to diminished classification.
The detailed list of selected summary statistics and their definitions can be found in Appendix~\ref{PPparams}.

%try fast and simple decision tree based ML algorithms. We find XGBoost Classifier to be the most suitable choice of ML algorithm to classify between a GW signal and glitch in the cWB framework. 

%XGBoost Classifier is a decision tree based ensemble learning algorithm \cite{XGBoost} that provides us with a continuous ranking criteria \Pxgb for each event.
 %The XGBoost model is trained based on gradient boosting, and the final result is the weighted average of the output obtained from each decision tree in the ensemble. 
%The output of a decision tree is given by the `leaf score' (estimated by the Log-Odds output) which helps in determining the category of a given observation. In XGBoost Classifier, the leaf scores from all the decision trees in the ensemble are added together to get a margin score. 

\subsection{Tuning XGBoost hyper-parameters} \label{sec:tuning}

XGBoost has a number of free hyper-parameters which control various properties of the learning process to prevent overfitting.
These hyper-parameters need to be tuned for each specific application.
%The \texttt{n\_estimators} parameter determines the total number of generated trees.
%The \texttt{learning\_rate} parameter regulates how much each generated tree effects the final prediction. The \texttt{max\_depth} parameter determines the deepness of each tree and thus effects the overall complexity of the model.
%Other parameters (\texttt{min\_child\_weight}, \texttt{colsample\_bytree}, \texttt{subsample}, \texttt{gamma}) act as different forms of regularization to preserve the conservative-ness of the algorithm and prevent overfitting. 
We use a small data set consisting of $200\, $yr of background data (which equates to approximately 500,000 noise events) and 2,000 simulated IMBH binary events to tune the XGBoost hyper-parameters. These background and simulation events are drawn from the O2 run.

%The data has two distinct classes; the background events and the simulation events. We use the multidimensional summary statistics output data of candidate events instead of time-series data for tuning, training and testing the XGBoost Classifier model. Tuning is the process of finding the optimal values for the hyper-parameters of the ML algorithm, which is performed based on a criteria of optimality (metric).

We perform a grid search over a range of six standard XGBoost hyper-parameters, listed in Table~\ref{tab:1}. We find the most optimal set by evaluating each configuration of XGBoost hyper-parameters with respect to the precision-recall area under the curve (AUC PR) metric~\cite{Davis2006} over 10-fold cross-validation. 
The optimal configuration of hyper-parameters, according to this criteria, is shown in bold in Table~\ref{tab:1}.
We use a method known as \textit{early stopping} to optimize the total number of trees generated. In early stopping, a small fraction of the training data set is set aside for validation. When the validation AUC PR score stops improving, the training ends to prevent the XGBoost from overfitting.
 
Overall, we found that the performance of the model was not highly dependent on the chosen hyper-parameter values. As a result, we keep this hyper-parameter configuration for all models presented in this paper.

\begin{table}[]
    \centering
    \begin{tabular}{cc}
        \hline
        \hline
        XGBoost hyper-parameter & entry \\
        \hline
        \hline
        \texttt{objective} & \texttt{binary:logistic}  \\
        \texttt{tree\_method} & \texttt{hist} \\
        \texttt{grow\_policy} & \texttt{lossguide} \\
        \texttt{n\_estimators} & 20,000$\dagger$ \\
        \texttt{max\_depth} & 7, 9, \textbf{11}, 13 \\
        \texttt{learning\_rate} & 0.1, \textbf{0.03} \\
        \texttt{min\_child\_weight} & 5.0, \textbf{10.0} \\
        \texttt{colsample\_bytree} & \textbf{0.6}, 0.8, 1.0 \\
        \texttt{subsample} & 0.4, 0.6, \textbf{0.8} \\
        \texttt{gamma} & 2.0, \textbf{5.0}, 10.0 \\
        \hline
        \hline
    \end{tabular}
    \caption{Entries for XGBoost hyper-parameters. We perform a grid search over 432 combinations of hyper-parameters. Bold entries indicate the optimal choice. $\dagger$: \texttt{n\_estimators} is optimized using early stopping, described in the main text.}
    \label{tab:1}
\end{table}

\subsection{XGBoost model training}
%\MSe{What's the difference between Tuning and Training?}{}

We train a separate ML model for each search configuration and each observing run. In this paper, we have two search configurations (BBH, IMBH) and analyze two observing runs (O1, O2), a total of four models.
The estimated central frequency $f_0$ of a GW signal is expected to be inversely proportional to the red-shifted total mass of the binary system.
As such, we select events with $60\, \mathrm{Hz} < f_0 < 300\, \mathrm{Hz}$ to train the BBH search models and events with $f_0 < 200\, \mathrm{Hz}$ to train the IMBH search models.
Each trained model uses the same XGBoost hyper-parameters discussed in Section~\ref{sec:tuning}, and consists of around $600$ total trees for O1 run and $1000$ total trees for O2 run, with an average of 20 leaves per tree.
%In the O3a run~\cite{GWTC2}, the IMBH configuration operates in the lower central frequency band $(24\, \mathrm{Hz} < f_0 < 100\, \mathrm{Hz})$ whereas the BBH configuration operates in the higher central frequency band $(60\, \mathrm{Hz} < f_0 < 300\, \mathrm{Hz})$.

For training, we select $100\, \mathrm{yr}$ of background data (approximately 250,000 noise events) per data chunk. For every $100\, \mathrm{yr}$ of background data we select approximately $1000$ simulation events.
We use simulated quasi-circular stellar-mass BBH mergers (simulation set i) to train the BBH search model, and we use simulated quasi-circular IMBH binary mergers (simulation set ii) for the IMBH search model.
The remaining background and simulation data are used for testing.
Generally, ML classifiers are expected to be more accurate when the same number of events per class are used for training.
However, in our case, using balanced classes is not feasible.
It is difficult to arbitrarily increase the number of simulated events due to the computational cost. 
Moreover, down-sampling the noise event set is not prudent since we could lose valuable information related to the high SNR tail of the background distribution.
Instead, we apply a weight to every sample noise event to reduce the class imbalance. This weight is dependent on $\eta_\mathrm{0}$ and gives less importance to the low SNR glitches. The weighting procedure is described in more detail in Appendix~\ref{sample_weight}.

%In order to adequately train the XGBoost model, we combine event data from all chunks in an observing run. We have 9 chunks of data in the first observing run O1, and 21 chunks of data in the second observing run O2. 
%For the O2 BBH configuration and O2 \texttt{IMBH configuration}, we select 8,625.4 years of background data for testing the XGBoost model. For the O1 BBH configuration we have 15,315.9 years and for the O1 IMBH configuration we have 3,132.3 years of background data for testing.

%Once the XGBoost model is trained, it is stored and used for estimating \Pxgb on the testing data as part of final analysis. XGBoost Classifier is found to perform well despite high class imbalance between simulation and background. Since the background distribution is different for different observing runs, we apply a $\eta_0$ dependent \textit{sample weight} to achieve a similar weighted background distribution for each observing run while avoiding high class imbalance (explained in Appendix~\ref{sample_weight}).

%We look at the necessary modifications to the existing cWB detection statistic to incorporate the continuous ML ranking criteria and improve the overall search efficiency.  

\subsection{cWB+ML detection statistic\label{det:sec}}

We incorporate the predictions made by the ML model directly into the detection statistic to improve noise rejection.
%First, we modify the main detection statistic used in the generic cWB search $\eta_0$ (Equation~\ref{eq:1}) to be more sensitive to the $\chi^2$ correction:
% \begin{equation}\label{eq:4}
% \eta_0 = \sqrt{\frac{E_\mathrm{c}}{1+\chi^2(\text{max}(1, \chi^2)-1)}}.
% \end{equation}
% Then, we define the reduced detection statistic used for the ML-enhanced cWB search as:
% \begin{equation}\label{eq:3}
% \eta_\mathrm{r} = \eta_0\, W_{\mathrm{XGB}}, 
% \end{equation}
We define the reduced detection statistic used for the ML-enhanced cWB search as:
\begin{equation}\label{eq:3}
\eta_\mathrm{r} = \sqrt{\frac{E_\mathrm{c}}{1+\chi^2(\text{max}(1, \chi^2)-1)}}\, W_{\mathrm{XGB}}, 
\end{equation}
where $W_{\mathrm{XGB}}$ is the XGBoost penalty factor. 
To compute $W_{\mathrm{XGB}}$, we first apply a correction to the XGBoost output, defined in Appendix~\ref{correction}. This correction is designed to suppress numerous noise events which have less than one cycle in the time domain waveform, which is typical for the known family of glitches found in the GW detector data.
Next, we apply a monotonic transformation, defined in Appendix~\ref{WXGB}, to obtain the penalty factor $W_{\mathrm{XGB}} = W_{\mathrm{XGB}}(P_{\mathrm{XGB}})$.
This transformation accentuates the ranking of events with the values of $P_{\mathrm{XGB}}$ very close to unity.

Although $W_{\mathrm{XGB}}$ itself could be used as a detection statistic, we find that it is susceptible to assigning high significance to low SNR noise events.
Instead, we use it as a penalty factor to the estimated effective correlated SNR $\eta_\mathrm{0}$.
The end result is a detection statistic $\eta_\mathrm{r}$ which is enhanced by the ML classification and resistant to overfitting the low SNR noise events. 

%This reduced detection statistic $\eta_\mathrm{r}$ is used for re-analysis of the O1 and O2 data in the cWB framework. 
%%%%%%%%%%%%%%%%%%%%%%%%%%%%%%%%%%%%%%%%%%%%%%%%

%%%%%%%%%%%%%%%%%%%%%%%%%%%%%%%%%%%%%%%%%%%%%%%%
\section{Results\label{res:sec}}
%\MSe{We have BBH and IMBH searches. Why are not they combined? I think it should be included, at least clarifying how to combine these two searches and how the trial factors are used.}{}

In this section, we present the results of the ML-enhanced cWB search and compare its sensitivity to the standard cWB search.
The significance of a given candidate event is estimated by its false alarm rate (FAR).
It is computed by counting the number of background events with the equal or higher value of the detection statistic than for the candidate event, divided by the total background time. 
\begin{figure}[h!]
    \centering
    \includegraphics[width=\linewidth]{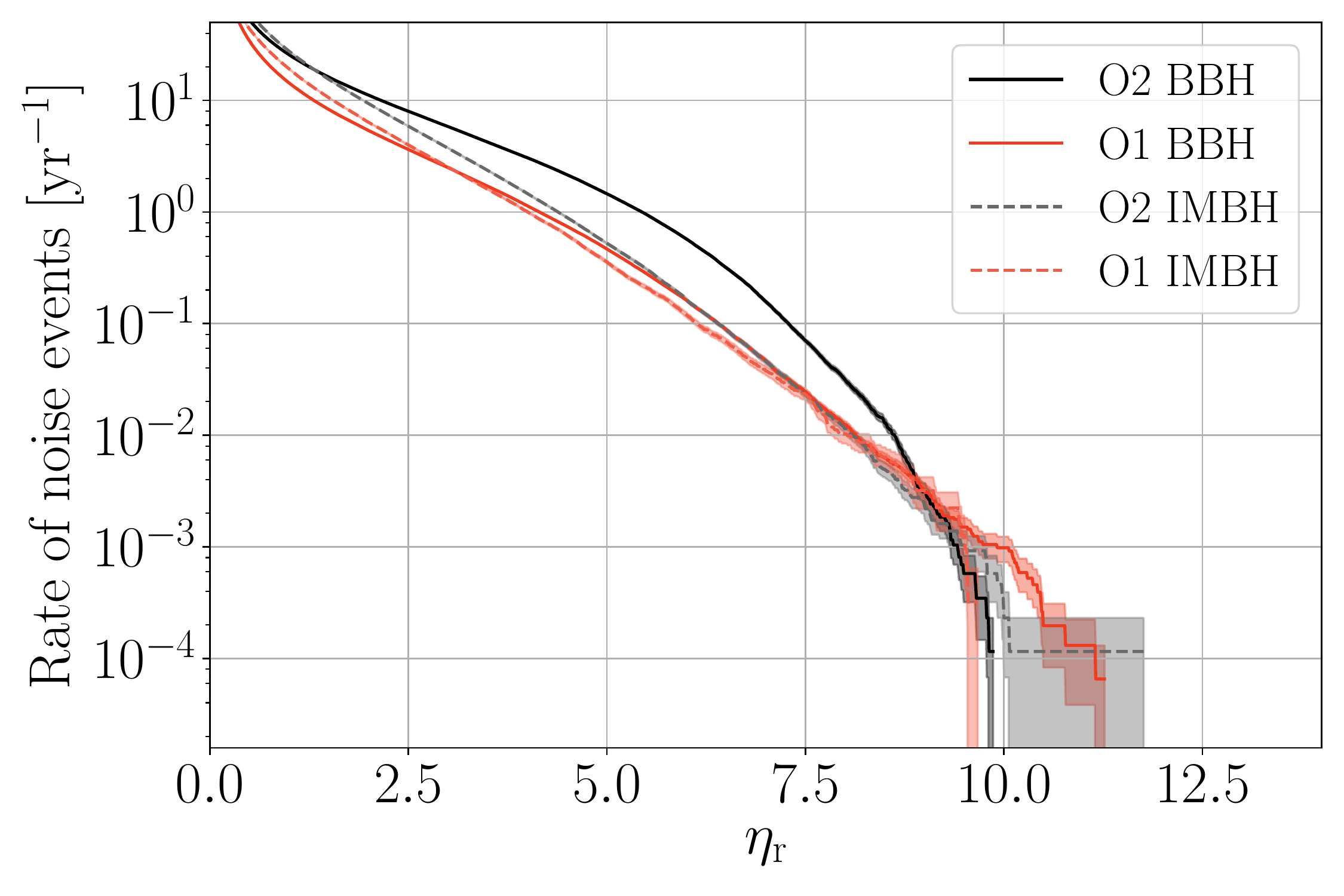}
    \caption{Rate of noise events vs detection statistic $\eta_\mathrm{r}$ for each search configuration (BBH, IMBH) and each observing run (O1, O2). We do not observe significant tails in the $\eta_\mathrm{r}$ distributions.}
    \label{fig:rate}
\end{figure}
\begin{figure*}[th]
    \centering
    \includegraphics[width=\textwidth]{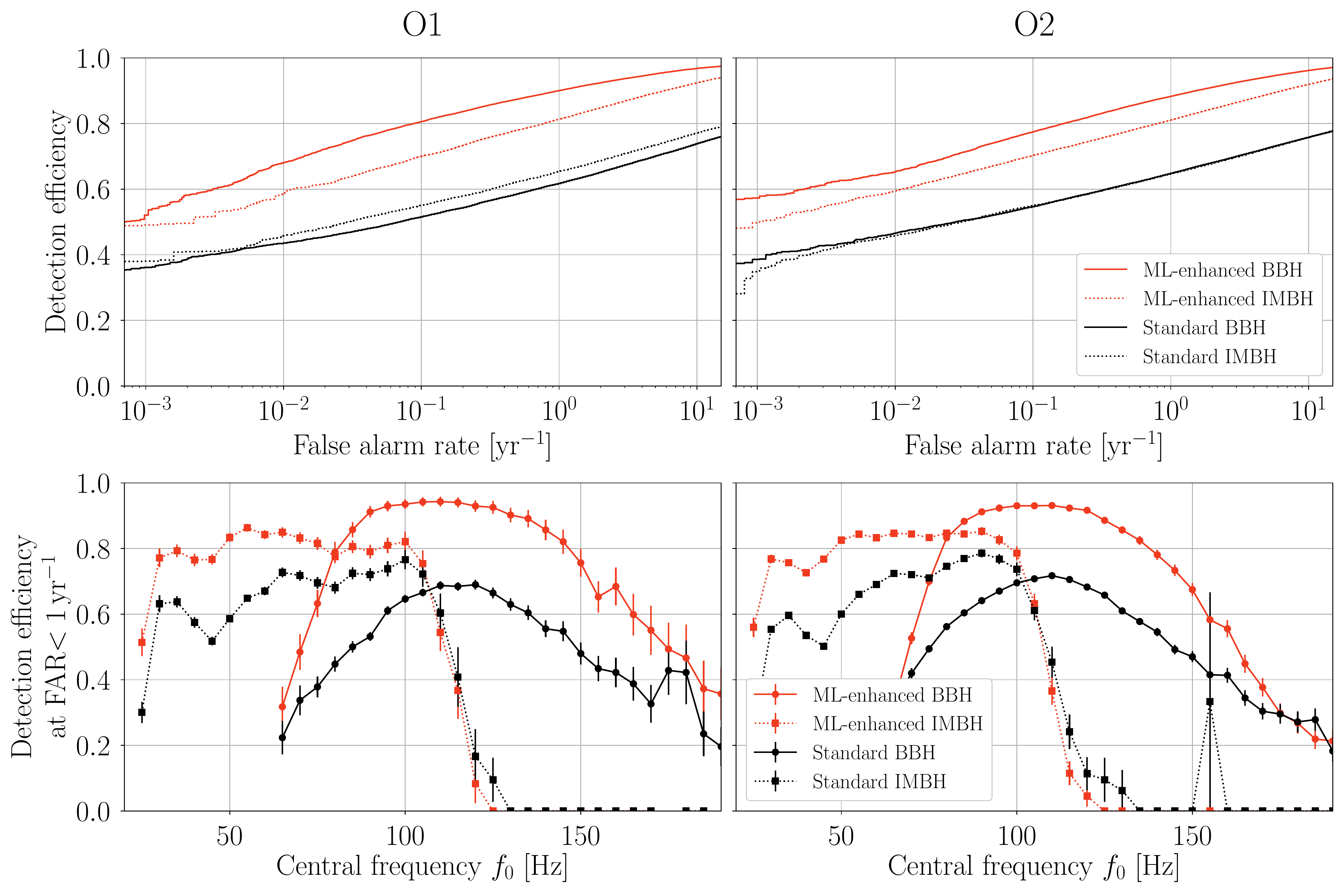}
    \caption{Top: Detection efficiency vs FAR for the O1 observing run (left) and the O2 observing run (right). Bottom: Detection efficiency for events identified with FAR less than $1\,$yr$^{-1}$ as a function of the central frequency $f_0$ for O1 (left) and O2 (right). Solid lines correspond to the BBH configuration, while dotted lines represent the IMBH configuration. Red curves represent the sensitivity of the ML-enhanced cWB search, and black curves represent the sensitivity of the standard cWB search with the veto method.}%, and green curves represent the sensitivity without the veto method.} %We observe an average improvement of $25.8\%$ of the detection efficiency at FAR $<$ 1 yr$^{-1}$, for the combined O1 and O2 \texttt{BBH configuration}, and an average improvement  of $16.8\%$  for the combined O1 and O2 \texttt{IMBH configuration}.}
    \label{fig:det_eff_o1o2}
\end{figure*}

For the ML-enhanced cWB search, FAR is calculated by using the reduced detection statistic $\eta_\mathrm{r}$  (Equation~\ref{eq:3}). Whereas for the standard cWB BBH search, we first remove vetoed events and then calculate FAR by using the $\eta_\mathrm{1}$ detection statistic (Equation~\ref{eqn2}). We compare the performance of the different cWB searches by comparing their detection efficiency defined
%searches estimate the sensitivity of the searches to different simulation sets with the detection efficiency defined 
%In this paper, we calculate the detection efficiency 
as the number of detected simulated events with FAR equal to or less than a given threshold, divided by the total number of recovered simulation events.

\subsection{Re-analysis of O1 and O2 data\label{res2:sec}}

First, we examine the background noise distributions for the ML-enhanced search.
Figure~\ref{fig:rate} shows the rate of the noise events for each search (BBH, IMBH) and each observing run (O1, O2) as a function of the reduced detection statistic $\eta_\mathrm{r}$. In all cases, we do not observe significant tails of the high $\eta_\mathrm{r}$ events, which indicates that the ML-enhanced detection statistic efficiently suppresses the high SNR outliers.

Next, we examine the sensitivities of each search for various simulation data sets.
The top panel of Figure~\ref{fig:det_eff_o1o2} shows the detection efficiency as a function of FAR for the various searches over the O1 (left) and O2 (right) data.
The ML-enhanced search (shown in red) is more sensitive than the standard search with vetoes (black) in the wide range of the FAR thresholds.
For events detected with FAR $<1\, \mathrm{yr}^{-1}$, we estimate the 26\% improvement for the BBH configuration and the 16\% improvement for the IMBH configuration, averaged over the two observing runs. For high significance detection (FAR $<100\, \mathrm{yr}^{-1}$), we estimate the 22\% improvement for the BBH configuration and the 13\% improvement for the IMBH configuration.

The bottom panel of Figure~\ref{fig:det_eff_o1o2} shows the detection efficiency for events detected with FAR $<1\, \mathrm{yr}^{-1}$ as a function of the central frequency $f_0$. Here, we see that for most frequency bins, the ML-enhanced search (red) is more sensitive than the standard search (black). This indicates that the ML-enhanced search is not overly tuned to any specific frequency bins but shows consistent improvement over the entire frequency range considered by the cWB search configurations. Since the central frequency $f_0$, is expected to be inversely proportional to the detector frame mass, we can infer that the ML-enhanced search is more sensitive over the entire BBH mass range accessible by LIGO.

Table~\ref{tab:2} reports the BBH candidates identified by the ML-enhanced cWB search in the O1-O2 observing runs. We recover 7 BBH candidates previously reported by the standard cWB search~\cite{GWTC1}, all identified with higher significance. Additionally, the ML-enhanced cWB search detected GW170809 with a FAR of $0.29\,\mathrm{yr}^{-1}$, which was previously vetoed in the standard cWB search. No other candidate events were identified with FAR less than $1\,\mathrm{yr}^{-1}$.

\begin{table}[bht]
    \centering
    \setlength{\tabcolsep}{6pt}
    
    \begin{tabular}{crr}
        \hline
        \hline
        Event & \multicolumn{1}{c}{Standard cWB} & \multicolumn{1}{c}{ ML-enhanced cWB} \\
         & \multicolumn{1}{c}{($\eta_\mathrm{1}$ + vetoes)} & \multicolumn{1}{c}{($\eta_{r}$)} \\
         & \multicolumn{1}{c}{FAR [yr$^{-1}$]} & \multicolumn{1}{c}{FAR [yr$^{-1}$]} \\
        \hline
        \hline
        GW150914 & $< 1.6 \times 10^{-4}$ & $< 7 \times 10^{-5}$ \\
        %\textcolor{red}{GW151012} & $...$ & $...$ \\
        GW151226 & $2 \times 10^{-2}$ & $6.5 \times 10^{-3}$ \\
        GW170104 & $2.9 \times 10^{-4}$ & $< 1.2 \times 10^{-4}$ \\
        GW170608 & $1.4 \times 10^{-4}$ & $< 1.2 \times 10^{-4}$ \\
        GW170729 & $2 \times 10^{-2}$ & $< 1.2 \times 10^{-4}$ \\
        GW170809 & \multicolumn{1}{c}{\,$\quad -$} & $2.9 \times 10^{-1}$ \\
        GW170814 & $< 2.1 \times 10^{-4}$ & $< 1.2 \times 10^{-4}$ \\
        %\textcolor{red}{GW170818} & $...$ & $...$ \\
        GW170823 & $2.1 \times 10^{-3}$ & $< 1.2 \times 10^{-4}$ \\
        \hline
        \hline
    \end{tabular}
    \caption{O1-O2 event candidates detected with the standard cWB BBH search ($\eta_\mathrm{1}$ + vetoes), and the ML-enhanced cWB BBH search ($\eta_\mathrm{r}$). We report all detections with FAR less than $1\, \mathrm{yr}^{-1}$. Entries with a `$<$' sign indicate that the estimated significance is limited by the amount of available background data.
    }
    \label{tab:2}
\end{table}

\subsection{Test of model robustness \label{res3:sec}}

\begin{figure}[h!]
    \centering
    \includegraphics[width=\linewidth]{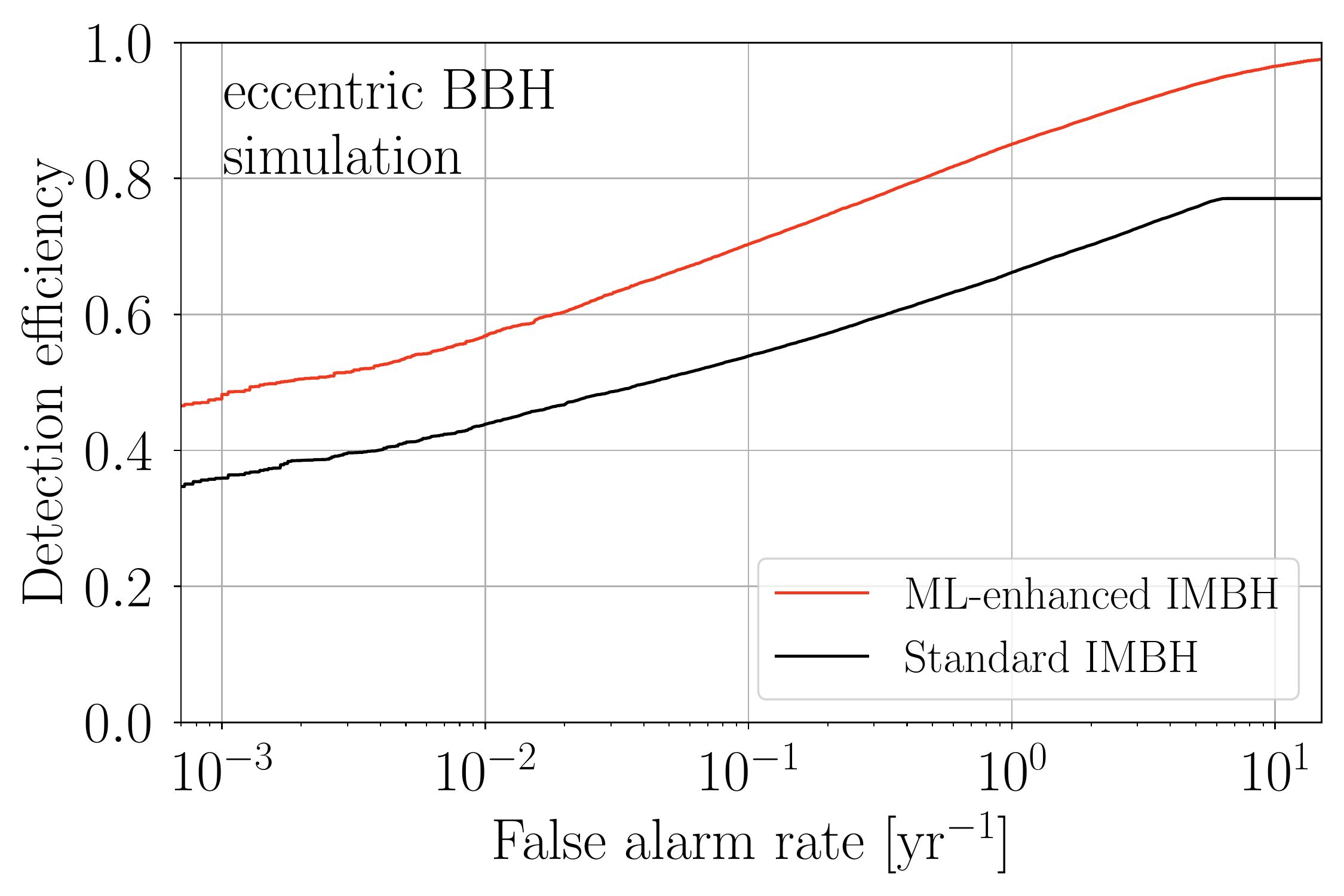}
    \caption{Detection efficiency vs FAR for a high mass, eccentric BBH simulation set recovered with the cWB IMBH configuration. The red curve represents the sensitivity of the ML-enhanced cWB search, and the black curve represents the sensitivity of the standard cWB search.}
    \label{fig:eBBH}
\end{figure}

As a final test, we analyze the performance of the ML-enhanced search on simulated waveforms outside of the training set. 
First, we investigate the sensitivity of the ML-enhanced IMBH search, which is trained on quasi-circular binaries,  to high mass BBH systems with highly eccentric orbits (simulation set iii).
%\textcolor{red}{\\ EBBH test dataset uses O2 Recolored to O3 SIM and BKG.\\ Precession test dataset uses O3A BBH precessing SIM and O3A BBH BKG ... Is this okay and should we explicitly mention this in the text/appendix?\\}
In Figure~\ref{fig:eBBH}, we show the sensitivity to the eccentric IMBH mergers of the ML-enhanced IMBH search compared to the standard IMBH search.
The ML-enhanced search is more sensitive to the eccentric BBH mergers despite the fact that we trained the ML model only on the quasi-circular IMBH waveforms.

\begin{figure}[h!]
    \centering
    \includegraphics[width=\linewidth]{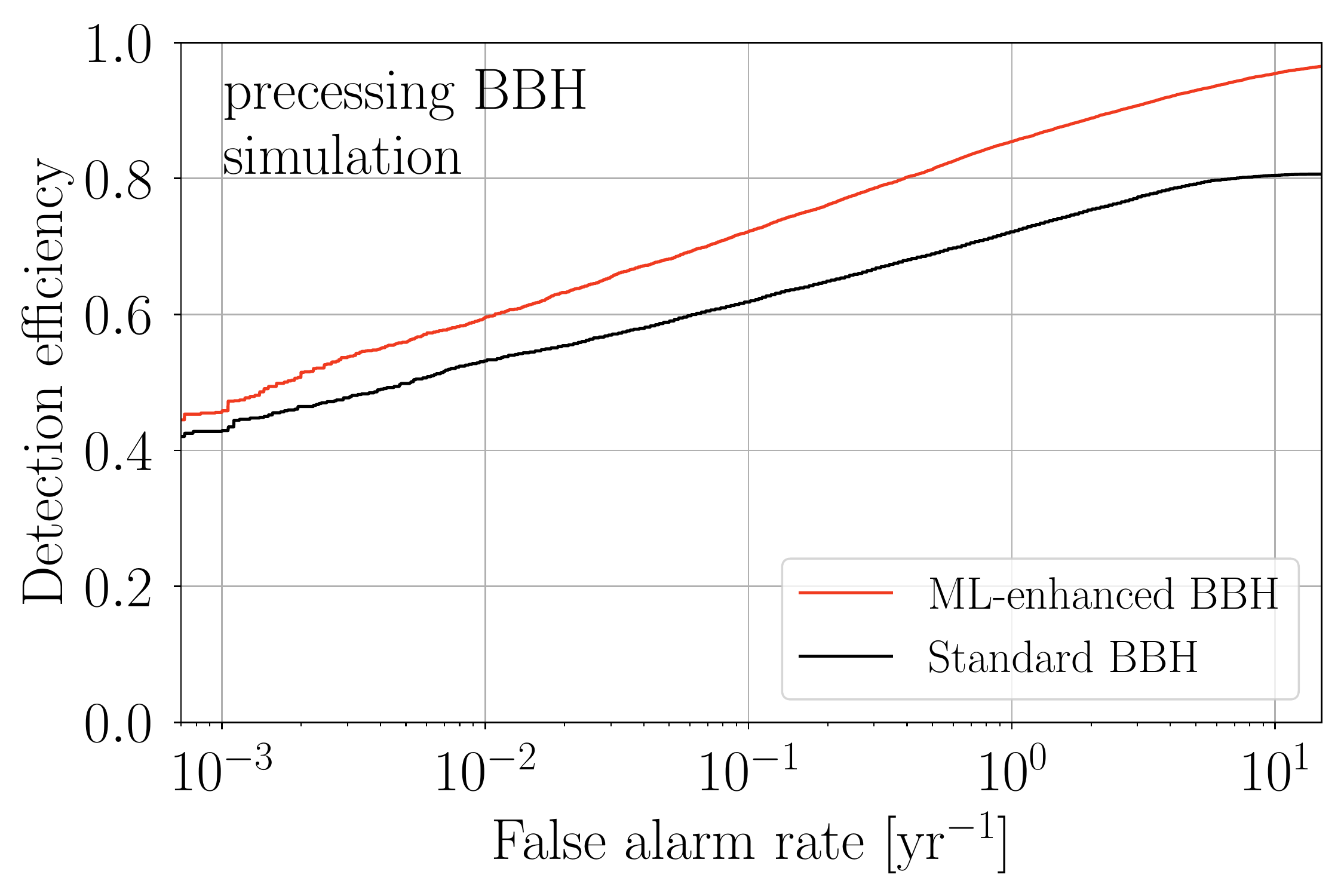}
    \caption{Detection efficiency vs FAR for a precessing BBH simulation set recovered with the cWB BBH configuration. The red curve represents the sensitivity of the ML-enhanced cWB search, and the black curve represents the sensitivity of the standard cWB search.}
    \label{fig:prec}
\end{figure}

Next, we test the ML-enhanced BBH search on precessing BBH systems (simulation set iv).
The training set consists of simulated waveforms with the (2,2) harmonic mode, aligned spins, and low mass ratio (1~-~1/4).
The testing set consists of simulated waveforms with higher-order modes, precessing spins, and a higher mass ratio (1~-~1/20).
In this case, the model is also trained on O2 data, whereas the testing set consists of O3a simulation and background, which has very different GW detector sensitivity. 
Figure~\ref{fig:prec} shows improved detection efficiency of the ML-enhanced BBH search compared to the standard BBH search. 

These results demonstrate that the ML-enhanced search is agnostic to the details of the BBH dynamical evolution including eccentricity, spin effects, and higher-order modes. It retains the robustness of the standard search, but with increased sensitivity.
%We train the XGBoost model on IMBH quasi-circular binaries from O2 data and test it on high mass BBH events with eccentric orbits. These eccentric BBH events are recovered by injecting signals with high eccentricity values (e = 0.66, 0.85, 0.94, 0.99) into O2 data rescaled to O3 sensitivity~\cite{szczepanczyk2021detecting}. 

\section{Conclusion} \label{conclusion:sec}

In conclusion, we introduce a novel method to automate the noise rejection in cWB with ML, and we demonstrate that the ML-enhanced cWB search has improved sensitivity to the BBH mergers.
Unlike the standard cWB search, which decimates noise events with the veto method, the ML-enhanced search uses a continuous ranking statistic and does not remove any events.
Instead, it penalizes noise events with the newly designed detection statistic $\eta_\mathrm{r}$, which combines the correlated network SNR with the penalty factor based on the signal-noise classification provided by the XGBoost model.
This new detection statistic $\eta_\mathrm{r}$ is resistant to loud noise glitches and allows cWB identification of BBH events with higher significance.

For the stellar-mass BBH mergers, the detection efficiency of the ML-enhanced cWB search is improved by 26\% compared to the standard cWB search (at FAR less than $1\, \mathrm{yr}^{-1}$). 
Similarly, for the simulation set of IMBH binary mergers, the detection efficiency is improved by 16\%.
While we do not claim any new BBH detections in the O1 and O2 data, we do improve the detection confidence of the previously identified GW candidates and recovered the GW170809 event missed by the standard cWB search.
%we estimate the FAR as $0.29\, \textrm{yr}^{-1}$.

The ML-enhanced cWB search is capable of detecting BBH signals well outside of the training set. We demonstrate the improved pipeline sensitivity to the highly eccentric BBH mergers despite only being trained on the quasi-circular IMBH signals. 
We also found the search to be agnostic to other binary waveform properties including precession, high mass ratio, and higher-order harmonic modes.
%Though  not  directly tested in this study,  we expect the search to be agnostic to the other binary waveform properties such as precession  and  higher  order  harmonic  modes,  not  directly included  in  the  training  data.
Waveforms with similar time-frequency morphology, but not predicted by general relativity should also be detectable.
%So, the cWB search retains its most valuable property: robustness.

The ML-enhanced BBH search is a promising addition to the cWB pipeline for future planned observing runs, where we expect numerous BBH detections. 
While in this study, we use only the LIGO Hanford and LIGO Livingston detector network, in future work we will expand the ML-enhanced search to the other detector networks which include Virgo and Kagra detectors. This will further improve the cWB sensitivity to the BBH mergers.
The ML-enhanced pipeline could be also used for the low latency cWB searches in the future observing runs. %to swiftly inform astronomers of potentially interesting BBH mergers.

\begin{acknowledgements}
This research has made use of data, software, and/or web tools obtained from the Gravitational Wave Open Science Center, a service of LIGO Laboratory, the LIGO Scientific Collaboration, and the Virgo Collaboration. This work was supported by the NSF Grant No. PHY 1806165. We gratefully acknowledge the support of LIGO and Virgo for the provision of computational resources. I.~B. acknowledges support by the NSF Grant No. PHY 1911796, the Alfred P. Sloan Foundation and by the University of Florida. 
We thank Ik Siong Heng, Erik Katsavounidis, Peter Shawhan, and Michele Zanolin for their continued participation and effort in reviewing cWB analyses.
We acknowledge the use of open source Python packages including \textsc{NumPy}~\cite{numpy}, \textsc{Pandas}~\cite{pandas}, \textsc{Matplotlib}~\cite{matplotlib}, and \textsc{scikit-learn}~\cite{scikit-learn}.
\end{acknowledgements}

%%%%%%%%%%%%%%%%%%%%%%%%%%%%%%%%%%%%%%%%%%%%%%%%

%%%%%%%%%%%%%%%%%%%%%%%%%%%%%%%%%%%%%%%%%%%%%%%%

\appendix
\section{cWB estimated summary statistics \label{PPparams}}
The ML algorithm is tuned and trained on the data from a selected subset of cWB summary statistics for each event. We start by truncating this subset by selecting the summary statistics that have a low correlation with each other. We also aggregate a few summary statistics together to prune the list of summary statistics that are used as input features for the ML algorithm. We select 14 cWB summary statistics in total that are used by the ML algorithm as the input list of features. The summary statistics are listed below:
\begin{itemize}
    \item $\eta_\mathrm{0}$ --- Main cWB detection statistic for the generic GW search. For the ML study, we cap the $\eta_\mathrm{0}$ value at 8 (any event with higher $\eta_\mathrm{0}$ is assigned a value of 8) so the algorithm is not affected by the high SNR events in the background distribution, which is a steep function of $\eta_\mathrm{0}$.
    \item $c_\mathrm{c}$ --- Coherent energy divided by the sum of coherent energy and null energy, defined in Ref.~\cite{Klimenko2008}.
    \item $n_\mathrm{f}$ --- Effective number of time-frequency resolutions used for event detection and waveform reconstruction.
    \item $E_\mathrm{c}/L$ --- Ratio of the coherent energy to the network likelihood.
    \item $\Delta T_\mathrm{s}$ --- Energy weighted signal duration.
    \item $\Delta F_\mathrm{s}$ --- Energy weighted signal bandwidth. 
    \item $f_0$ --- Energy weighted signal central frequency. 
    \item $\mathcal{M}$ --- Chirp mass parameter estimated in the time-frequency domain, defined in Ref.~\cite{Tiwari2015_chirp}.
    \item $e_{\mathrm{M}}$ --- Chirp mass goodness of the fit metric, presented in Ref.~\cite{Tiwari2015_chirp}. 
    \item $Q_\mathrm{a}$ --- The waveform shape parameter~\cite{Qveto} developed to identify a characteristic family of (blip) glitches present in the detectors~\cite{GWTC1, McIver2012}. A value of $Q_\mathrm{a}$ corresponds to the cWB event being a blip glitch.
    \item $Q_\mathrm{p}$ --- An estimation of the effective number of cycles in a cWB event. 
    %This is computed by dividing the quality factor of the reconstructed waveform by an appropriate function of the coherent energy.
    \item $L_\mathrm{v}$--- for the loudest pixel, the ratio between the pixel energy and the total energy of an event~\cite{Lveto2}.%Lveto2
    \item $\chi^2$ --- quality of the event reconstruction, $\chi^2 = E_\mathrm{n} / N_{\mathrm{df}}$ where $E_\mathrm{n}$ is the residual noise energy estimated and $N_{\mathrm{df}}$ is the number of independent wavelet amplitudes describing an event.
    \item $C_\mathrm{n}$ --- Data chunk number. LIGO-Virgo data is divided into time segments known as chunks, which typically contain a few days of strain data. Including the data chunk number allows the ML algorithm to respond to changes in detector sensitivity across separate observing runs and chunks.
    
    %\textbf{chunk number}: cWB pipeline analyses the data in different time intervals of observed data called chunks. We use chunk number as an input parameter so that the ML algorithm may be sensitive to the different background distribution in the chunks. 
\end{itemize} 
% The above list of cWB parameters serve as the feature list for the ML algorithm.

\section{Noise event sample weight\label{sample_weight}}

\begin{figure}[H]
    \centering
    \includegraphics[width=0.99\linewidth]{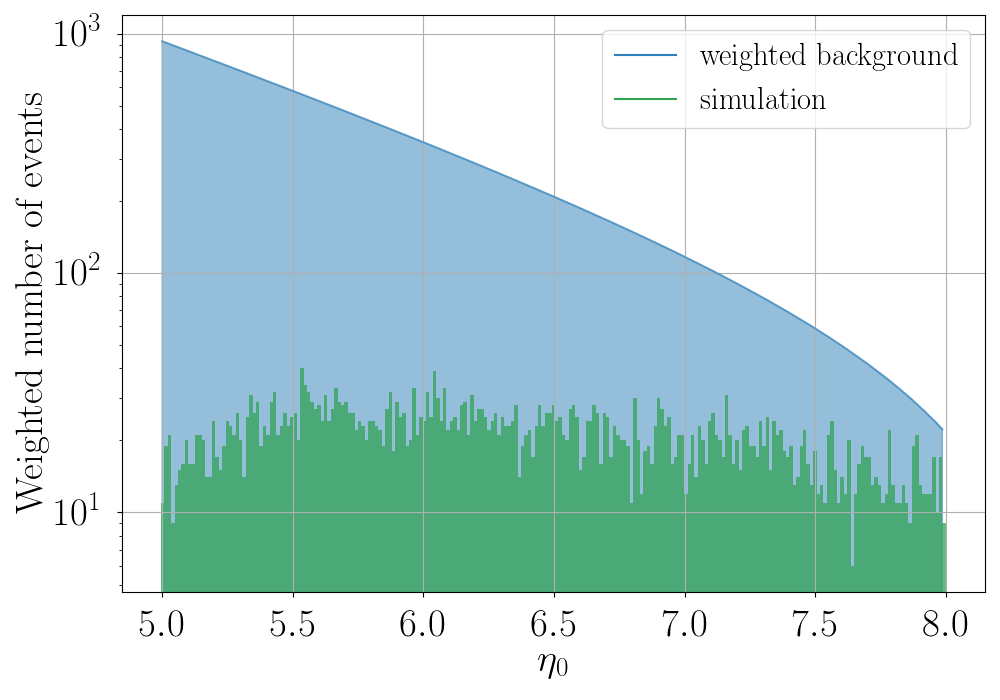}
    \caption{Weighted event distribution as a function of $\eta_\mathrm{0}$ for background events and simulated events. Here, we show the IMBH configuration with the O1 run data.}
    \label{fig:sample_weight}
\end{figure}

%The significance of the cWB events are calculated with respect to the rate of the background events.
In the initial testing phase with the trained XGBoost model, we found a tail of high SNR background events, most of which were consistent with blip glitches. This tail is caused by the suboptimal ML model due to the high class imbalance between the high SNR noise events and signal events.  To correct for this tail, we applied an $\eta_\mathrm{0}$ dependent sample weight to the noise events while training the XGBoost models. This weight provides us with a weighted background distribution as shown in Figure~\ref{fig:sample_weight}, which is similar for any observing run. 
The sample weight for the simulation events is set to 1. For the noise events, we divide $\eta_\mathrm{0}$ into 231 bins with values ranging from 5 to 8 and apply the sample weight $w_\mathrm{B}$ for each bin as follows:
\begin{equation}
% \text{BKG[bin]} = \frac{(\text{SIM$_{avg}$} - 65) + 0.8*\exp (0.9*(12.9-\hat{\eta}))}{\text{BKG[bin].}}
w_{\mathrm{B}}(\eta_\mathrm{0}) = \frac{(N_\mathrm{S}(\eta_\mathrm{0}) - a) + b\, e^{c\left(d-\eta_\mathrm{0}\right)} }{N_\mathrm{B}(\eta_\mathrm{0})} \, ,
\end{equation}
where $N_\mathrm{S}(\eta_\mathrm{0})$ is the number of signal events and $N_\mathrm{B}(\eta_\mathrm{0})$ is the number of noise events, in a given bin. We found that the sample weight works reasonably well for the following choices of values: $a = 65$, $b = 0.8$, $c = 0.9,$ and $d = 12.9$.
The number of simulation events for $\eta_\mathrm{0} \geq$ 8 were re-sampled to match the number of background events with the same range of $\eta_\mathrm{0}$. This application of sample weight enables the algorithm to successfully remove almost all the high SNR outliers while keeping the weighted events class imbalance ($N_\text{B}/N_\text{S}$) to around 20 for any given observing run. Without the application of the sample weight, the class imbalance ranges from 50 to 600 or even higher for the given training setup depending on the observing run and the search configuration.
\begin{figure}[H]
    \centering
    \includegraphics[width=0.99\linewidth]{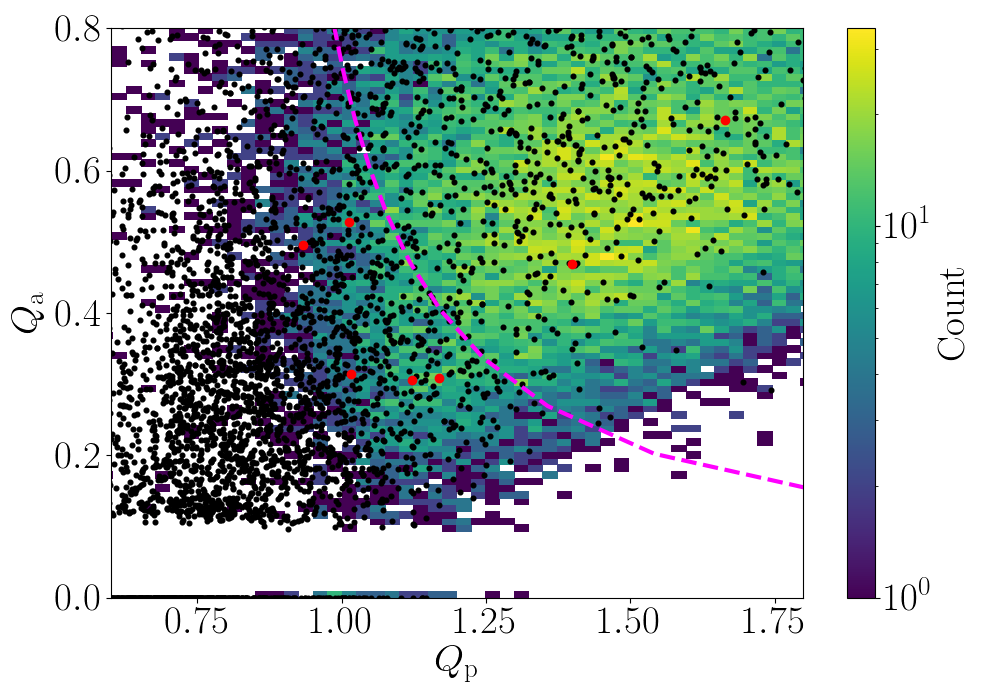}   
    \caption{$Q_\mathrm{a}-Q_\mathrm{p}$ parameter space plot for the IMBH configuration with O1 run data. Black dots represent loud background noise events ($\eta_\mathrm{0} \geq 8$). Red dots represent loudest background outliers ($\eta_\mathrm{0} \geq 10$). The heatmap represents the simulated IMBH distribution.}
    \label{fig:Qa_Qp}
\end{figure}

\section{XGBoost penalty factor}

\subsection{cWB correction to \Pxgb in the IMBH configuration\label{correction}}

Further investigation revealed that in the IMBH configuration, high SNR background outliers were found in a specific parameter space represented by the $Q_\mathrm{a}$-$Q_\mathrm{p}$ plot in Figure~\ref{fig:Qa_Qp}. In the standard veto method, the application of the $Q_\mathrm{a}$ and $Q_\mathrm{p}$ vetoes would have removed all events below the predefined thresholds at the cost of losing a small fraction of simulated events. In the ML method, we apply a correction to the ranking criteria \Pxgb such that we suppress the background outliers in the affected parameter space below the magenta-colored curve in Figure~\ref{fig:Qa_Qp}.  

The correction to the continuous ranking criteria output \Pxgb from XGBoost is applied only to the IMBH configuration (shorter signals with fewer cycles in the time domain), given by the following equations:
\begin{equation}
P_{\mathrm{XGB}}= \begin{cases} P_{\mathrm{XGB}} - (0.15 - Q_\mathrm{a}(Q_\mathrm{p}-0.8)),\\
\qquad \text{if } Q_\mathrm{a}(Q_\mathrm{p}-0.8) \leq 0.15 \\ \qquad \text{ (under the curve)} \\
P_{\mathrm{XGB}},\\
\qquad \text{if }Q_\mathrm{a}(Q_\mathrm{p}-0.8) > 0.15 \\ \qquad \text{ (above the curve)}.\end{cases}
\end{equation}

This correction enhances detection of signal and suppresses glitches in the desired part of the $Q_\mathrm{a}$-$Q_\mathrm{p}$ parameter space explicitly without any changes to the XGBoost training and testing procedure. The correction comes at the cost of losing a small fraction of IMBH signals with small values of $Q_\mathrm{p}$.

\subsection{Monotonic transformation $W_{\mathrm{XGB}}$} \label{WXGB}

The monotonic transformation applied on $P_{\mathrm{XGB}}$ is defined as follows,
\begin{equation}\label{eq:WXGB}
W_{\mathrm{XGB}}(P_{\mathrm{XGB}}) =   \frac{-\log (1.0 - 0.995\sqrt{P_{\mathrm{XGB}}})}{5.3},
\end{equation}
which counterbalances the steep sigmoid function used by XGBoost while producing $P_{\mathrm{XGB}}$. The $P_{\mathrm{XGB}}$ output by XGBoost for any event has a precision of up to 5 decimal places. As a result, the effect of using $P_{\mathrm{XGB}}$ directly as a penalty factor is not very significant. The transformation in Equation~\ref{eq:WXGB} allows us to zoom in on the high precision, for $P_{\mathrm{XGB}}$ values close to unity. This helps us differentiate between high SNR background events and simulation events that typically end up with very high values of $P_{\mathrm{XGB}}$.

\bibliography{main.bib}
\end{document}